\def\ARXIVVERSION{1}
\documentclass[11pt]{article}

\ifdefined\ARXIVVERSION
\usepackage[preprint]{acl}
\else
\usepackage[review]{acl}
\fi
\usepackage{times}
\usepackage{latexsym}
\usepackage{amsmath}
\usepackage{booktabs}
\usepackage{graphicx}
\usepackage{float}
\usepackage{placeins}
\usepackage[T1]{fontenc}
\usepackage[utf8]{inputenc}
\usepackage{microtype}
\usepackage{inconsolata}
\usepackage{url}

\newsavebox{\promptboxcontent}
\newenvironment{promptbox}[1]{%
  \par\smallskip
  \begingroup
  \def\promptboxtitle{#1}%
  \setlength{\fboxsep}{0pt}%
  \begin{lrbox}{\promptboxcontent}%
  \begin{minipage}{\dimexpr\linewidth-2\fboxrule\relax}
  \colorbox{black!55}{\makebox[\linewidth][l]{\hspace{6pt}%
    \color{white}\bfseries\small\strut\promptboxtitle}}%
  \par\vspace{6pt}
  \hspace*{6pt}\begin{minipage}{\dimexpr\linewidth-12pt\relax}
  \small\setlength{\parindent}{0pt}
}{%
  \end{minipage}\par\vspace{6pt}
  \end{minipage}%
  \end{lrbox}%
  \noindent\fcolorbox{black!55}{black!2}{\usebox{\promptboxcontent}}%
  \endgroup
  \par\smallskip
}

\title{Query Expansion Should Be Coordinated: Dense Expands, Sparse Anchors}

\ifdefined\ARXIVVERSION
\author{
  Chunran Zhang \\
  School of Computing and Artificial Intelligence \\
  Southwest Jiaotong University \\
  Chengdu, China \\
  \texttt{chronis@my.swjtu.edu.cn}
}
\else
\author{Anonymous ACL submission}
\fi

\begin{document}
\maketitle

\begin{abstract}
Retrieval-augmented generation (RAG) systems rely on retrieval modules to
ground large language model (LLM) outputs. LLM-based query expansion enriches
retrieval with document-like passages, but evaluations of hybrid retrieval
often fuse fixed top-$L$ prefixes of dense and sparse rankings. Because $L$
controls cross-channel contributions and ranking access, it can alter measured
expansion gains. We therefore evaluate complete-list effectiveness and record
per-channel replay stopping depths required to certify the ordered top-$K$.
This changes the design: because both rankings determine the fused result,
their query constructions should be coordinated rather than designed
independently.
We present DESA (Dense Expansion and Sparse Anchoring), which shares generated
references across channels but specializes their integration. Orthogonal
residual expansion adds new semantic directions to the dense query, whereas
score-product anchoring reorders the original sparse support without admitting
expansion-only matches. The same references thus play complementary roles:
Dense expands; Sparse anchors. Across seven BEIR datasets, DESA improves
nDCG@10 and Recall@20 over the unexpanded query by 3.82\% and 2.38\%, while
reducing dense and sparse replay stopping depths by 36.90\% and 36.56\%.
\end{abstract}

\section{Introduction}
\label{sec:introduction}

LLM-based query expansion in hybrid retrieval is commonly evaluated by fusing
fixed top-$L$ prefixes from dense and sparse rankings
\citep{gao-etal-2023-precise,wang-etal-2023-query2doc,zhang-etal-2024-exploring-best,liu-zhang-2025-exp4fuse}.
Yet $L$ controls both ranking access and which cross-channel contributions
enter fusion, so changing it may change the fused result itself. We instead
evaluate effectiveness from the complete rankings, then replay them to record
the per-channel depths needed to certify the same ordered top-$K$. This
separates the result produced by an expansion from the ranked evidence required
to establish it. We ask whether query expansion can improve the former while
reducing both depths relative to the unexpanded query.

\begin{figure}[!t]
  \centering
  \includegraphics[width=\columnwidth]{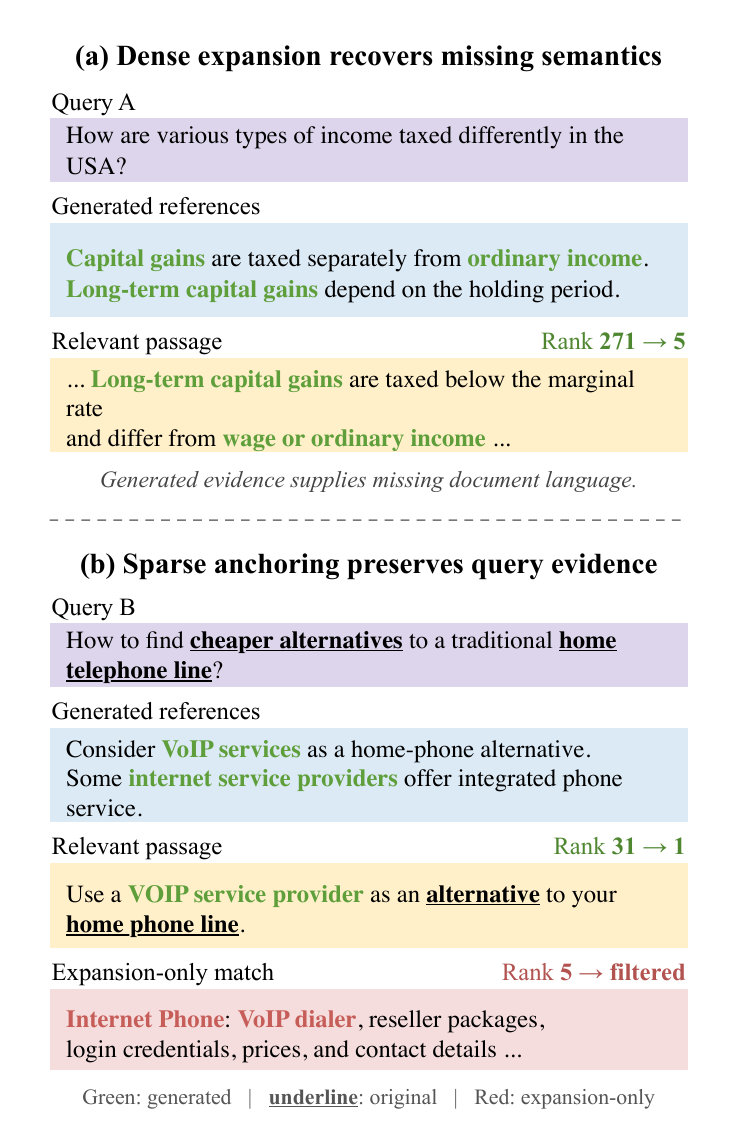}
  \caption{FiQA examples of DESA \citep{maia-etal-2018-fiqa,thakur-etal-2021-beir}.
  Dense expansion recovers missing document semantics; sparse anchoring promotes
  jointly supported passages and excludes expansion-only matches. Two references
  are shown per query; text is abridged and paraphrased, and ranks use draw~0.}
  \label{fig:desa-intuition}
\end{figure}

This evaluation view changes the design problem. The fused result and the
depths required to certify it depend jointly on the dense and sparse rankings,
so query construction for the two channels cannot be designed independently.
Retriever-specific rewriting adapts each query to its own retriever, but does
not explicitly control the cross-channel relation that governs exact fusion
\citep{li-etal-2026-dvcqr}. We instead couple the two constructions through one
shared set of generated references. Shared generation alone, however, is
insufficient: applied directly to both channels, it enlarges sparse support on
six of seven datasets, by as much as $5.72\times$, and on NFCorpus reduces dense
depth while increasing sparse depth. These observations motivate a coupled
design that shares the evidence basis across channels but constrains its role
within each channel.

DESA (Dense Expansion and Sparse Anchoring) implements this shared-generation,
channel-specific-integration design. The LLM generates complementary reference
passages once and supplies the same references to both channels. In dense
retrieval, orthogonal residual expansion preserves the original query direction
while adding the new semantic component supplied by the references. In sparse
retrieval, score-product anchoring allows the references to reorder originally
supported documents without introducing expansion-only matches. Together,
these operators coordinate the two channels through shared evidence: dense
retrieval can acquire new semantic support, while sparse retrieval remains
anchored to the original query support (Figure~\ref{fig:desa-intuition}).

Across seven BEIR datasets, DESA improves equal-dataset macro
nDCG@10 and Recall@20 over the unexpanded query by 3.82\% and 2.38\%, while
reducing dense and sparse depths by 36.90\% and 36.56\%, respectively. With
equal dataset weighting, 63.31\% of queries become shallower in both channels.
Compared with matched Shared expansion, DESA significantly improves nDCG while
preserving the original sparse support. Component analyses identify sparse
anchoring as the more consistent source of effectiveness gains, with dense
expansion providing complementary gains across most datasets.

\section{Related Work}
\label{sec:related-work}

\subsection{Evaluating Expansion under Hybrid Fusion}

Hybrid query-expansion studies commonly evaluate fused rankings constructed
from fixed top-$L$ prefixes
\citep{zhang-etal-2024-exploring-best,liu-zhang-2025-exp4fuse}. Here, $L$
controls both how much of each ranking is accessed and which cross-channel
contributions enter fusion, making the measured expansion effect conditional
on the selected cutoff. We therefore separate the complete-list fusion target
from the per-channel replay depths required to certify it. Applied to query
expansion, this distinction reveals how query construction changes both the
fused result and its certification depths.

\subsection{LLM-Based Query Expansion}

LLMs can construct document-like expansions containing lexical and semantic
evidence absent from the original query. HyDE encodes a generated hypothetical
document for dense retrieval, whereas Query2doc appends a pseudo-document to
the query for sparse and dense retrieval
\citep{gao-etal-2023-precise,wang-etal-2023-query2doc}. Later methods generate
different forms of evidence or sample multiple references to cover alternative
expressions of the same information need
\citep{jagerman-etal-2023-query-expansion,zhang-etal-2024-exploring-best}.
Although generated text can provide useful retrieval cues, its effects vary
across queries, datasets, and retrievers, and irrelevant content may reduce
effectiveness \citep{weller-etal-2024-generative}.

\subsection{Retriever-Aware Expansion and Integration}

Prior work addresses retriever dependence at three stages: generation,
integration, and fusion. DVCQR generates separate lexical and semantic rewrites
for sparse and dense retrieval \citep{li-etal-2026-dvcqr}. At integration, MuGI
applies different formulas to shared pseudo-references for the two retrievers,
while Word2Passage reweights generated terms to reduce lexical noise
\citep{zhang-etal-2024-exploring-best,choi-etal-2025-word2passage}. At fusion,
Exp4Fuse combines original and expanded rankings, whereas QuDAR adaptively
combines four routes formed by original and expanded queries under both
retriever types \citep{liu-zhang-2025-exp4fuse,kim-etal-2026-qudar}.

DESA departs from this channel-wise view. Because complete-list effectiveness
and per-channel certification depths are properties of the fused result, the
two channels cannot be designed independently. Prior methods adapt generation,
integration, or fusion to retriever-specific behavior, but do not use this
cross-channel dependence as a design constraint. DESA instead couples dense
and sparse retrieval under an effectiveness--depth objective, sharing evidence
across channels to add dense semantic directions while only reordering the
original sparse support.

\section{DESA}
\label{sec:method}

\subsection{Coupled Construction Overview}

\begin{figure}[H]
  \centering
  \includegraphics[width=\columnwidth]{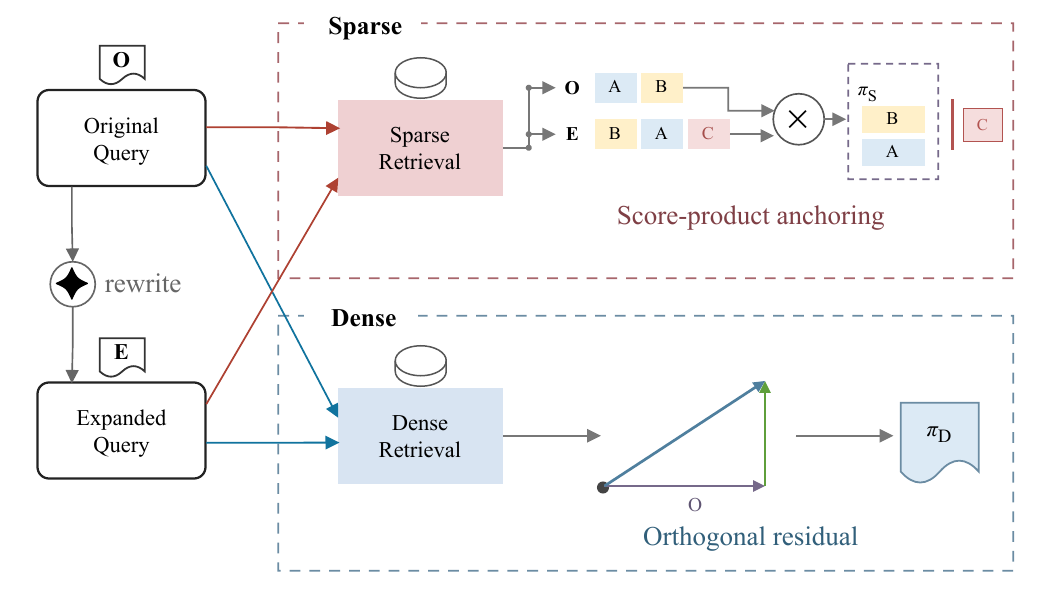}
  \caption{Coupled construction in DESA. A shared reference set establishes a
  common evidence basis. Orthogonal residual expansion adds new dense semantic
  directions, whereas score-product anchoring restricts sparse reordering to
  the original query's support.}
  \label{fig:desa-overview}
\end{figure}

DESA jointly designs dense and sparse query construction for the fused result,
aiming to improve complete-list effectiveness while reducing both certification
depths relative to the unexpanded query. Given an original query $q_o$, DESA
first generates one set of complementary reference passages
$R=\{r_1,\ldots,r_n\}$ shared by both channels. This common evidence basis
prevents channel differences from originating in independently generated
rewrites, placing specialization entirely in integration. Dense retrieval may
incorporate new semantic directions from the references, whereas sparse
retrieval may only reorder documents supported by the original query. The
resulting dense and sparse rankings are fused into the final result.
Figure~\ref{fig:desa-overview} summarizes this coupled construction.

\subsection{Shared Reference Generation}

DESA generates one reference set before constructing either channel ranking,
rather than producing a separate rewrite for each retriever. The generated
evidence is therefore held common across channels, so their differences arise
from controlled integration rather than different generated content.

The generator is prompted to produce $n$ short passages that could plausibly
occur in relevant documents. Each passage introduces a different alias,
technical term, or paraphrase while retaining the entities, relations,
numbers, polarity, temporal conditions, scope, and other constraints of
$q_o$. The prompt prohibits broader information needs and invented answers.
The passages therefore vary in expression rather than intent. The same set
$R$ is supplied unchanged to both channel operators. The complete prompt and
decoding settings are provided in Appendix~A.

\subsection{Coupled Dense--Sparse Construction}

The two operators act on the same original query and generated references but
assign different roles to the shared evidence. Dense retrieval permits the
references to add semantic information, subject to bounded deviation from the
original query. Sparse retrieval permits them to alter ranking order, but not
to admit documents unsupported by the original query.

\subsubsection{Orthogonal Residual Expansion}

The dense branch should incorporate semantic information absent from the
original query without allowing the generated references to replace its
original direction. Let $f_D$ be the dense encoder. We encode the original
query and each query--reference pair as unit vectors:

\[
\begin{aligned}
\mathbf v_o&=\operatorname{norm}(f_D(q_o)),\\
\mathbf v_i&=\operatorname{norm}
\!\left(f_D(\operatorname{concat}(q_o,r_i))\right),\\
\bar{\mathbf v}_r&=\frac{1}{n}\sum_{i=1}^{n}\mathbf v_i .
\end{aligned}
\]

We remove the component of $\bar{\mathbf v}_r$ parallel to the original query
and add the remaining residual to $\mathbf v_o$:

\[
\mathbf z=
\bar{\mathbf v}_r-
(\mathbf v_o^\top\bar{\mathbf v}_r)\mathbf v_o,
\qquad
\mathbf e_D=\operatorname{norm}(\mathbf v_o+\mathbf z).
\]

Documents are ranked by their similarity to $\mathbf e_D$, producing the
dense ranking $\pi_D$. Because $\mathbf z\perp\mathbf v_o$ and
$\|\mathbf z\|\leq1$,

\[
\angle(\mathbf e_D,\mathbf v_o)
=\arctan\|\mathbf z\|
\leq\frac{\pi}{4}.
\]

The references can therefore add semantic directions absent from the original
representation while keeping the expanded query within $45^\circ$ of it. If
they contribute no orthogonal component, then $\mathbf e_D=\mathbf v_o$.
Orthogonal residual expansion thus grants the shared evidence additive
semantic authority through a bounded, parameter-free update.

\subsubsection{Score-Product Anchoring}

The sparse branch requires a different constraint. Directly adding generated
terms may admit expansion-only documents, changing the sparse support and its
relation to the dense ranking. DESA instead allows generated lexical evidence
to reorder documents only within the original sparse support.

The expanded query concatenates the original query and all reference passages:

\[
q_r=\operatorname{concat}(q_o,r_1,\ldots,r_n).
\]

Let $s_S(q,d)$ be a non-negative additive sparse scorer. Each document $d$
receives an original score, an expanded score, and their product:

\[
\begin{aligned}
s_o(d)&=s_S(q_o,d), &
s_r(d)&=s_S(q_r,d),\\
s_A(d)&=s_o(d)s_r(d).&&
\end{aligned}
\]

Because $q_r$ contains $q_o$, every document supported by the original query
is also supported by the expanded query. With
$\operatorname{supp}(s)=\{d:s(d)>0\}$,

\[
\operatorname{supp}(s_A)
=\operatorname{supp}(s_o)\cap\operatorname{supp}(s_r)
=\operatorname{supp}(s_o).
\]

Sorting the positive anchored scores produces the sparse ranking $\pi_S$.
Generated terms may promote or demote documents within the original support,
but documents matched only by the expansion receive zero anchored score and
are excluded. If $s_r$ is merely a positive multiple of $s_o$, the original
sparse order is unchanged; positive global score rescaling also leaves the
ranking invariant. Score-product anchoring therefore grants the shared
evidence reordering authority but no independent sparse admission authority.

\section{Experiments}
\label{sec:experiments}

We test cutoff sensitivity, the joint effectiveness--depth outcome, the role
of coupled construction, and robustness across model and corpus choices.

\subsection{Evaluation Protocol}
\label{sec:quality-access}

Each method is evaluated on its complete-list fusion target. For two-channel
methods, complete dense and sparse rankings are fused by equal-weight RRF
\citep{cormack-etal-2009-rrf}:

\[
F(d)=
\sum_{c\in\{D,S\}}
\frac{1}{\kappa+\operatorname{rank}_{\pi_c}(d)}.
\]

A document absent from a channel contributes zero. Sorting by $F(d)$, with
document identifiers breaking ties, defines the ordered target $T_K$. We set
$K=20$ and evaluate effectiveness on this complete-list result.

Bound-based replay then records how deeply each ranking must be read to certify
$T_K$ \citep{zhang2026exact}. After revealing $L_c$ items from channel $c$, the
next unread contribution is bounded by

\[
u_c(L_c)=\frac{1}{\kappa+L_c+1}.
\]

Replay reads the channel with the larger bound, breaking ties toward Dense,
until no unread document can change the membership or order of $T_K$. The
stopping pair $(L_D,L_S)$ is the certification depth of that target.

The matched QuDAR control uses four rankings,
$c\in\{OS,OD,ES,ED\}$, and channel weights $w_c$:

\[
\begin{aligned}
F(d) &= \sum_c \frac{w_c}
{\kappa+\operatorname{rank}_{\pi_c}(d)}, \\
u_c(L_c) &= \frac{w_c}{\kappa+L_c+1}.
\end{aligned}
\]

For QuDAR-complete, $w_c=1/4$ across the four rankings; confidence-derived
weights are used only in the diagnostic.
We compare QuDAR's total depth, $\sum_c L_c$, with DESA's $L_D+L_S$.

The cutoff diagnostic instead fuses prefixes at
$L\in\{10,20,50,100,200,500,1000\}$. For each query and draw, we label the
change from Original as a gain, tie, or loss under fixed-$L$ and complete-list
fusion. We report label changes, strict gain--loss reversals, and agreement with
$T_{20}$.

All other experiments report nDCG@10 and Recall@20
\citep{jarvelin-kekalainen-2002-cumulated}, $(L_D,L_S)$, and the fraction of
queries shallower in both channels. Sparse support and exhaustion diagnose
changes in lexical access.

\subsection{Datasets and Retrieval Setup}

We evaluate seven English BEIR datasets \citep{thakur-etal-2021-beir}: SciFact
\citep{wadden-etal-2020-fact}, NFCorpus \citep{boteva-etal-2016-fulltext},
TREC-COVID \citep{voorhees-etal-2020-trec-covid}, FiQA
\citep{maia-etal-2018-fiqa}, ArguAna \citep{wachsmuth-etal-2018-retrieval},
Touch\'e-2020 \citep{bondarenko-etal-2020-touche}, and SCIDOCS
\citep{cohan-etal-2020-specter}. They span scientific claim verification,
biomedical and financial retrieval, argument retrieval, and scientific
recommendation. Unless noted otherwise, all analyses use the seven datasets
with equal macro weight.

The scale analysis uses nested TREC-COVID corpora of 25,000, 50,000, 100,000,
and all documents. Each snapshot retains every judged relevant document and is
a strict subset of the next.

Qwen2.5-7B-Instruct is the primary generator
\citep{qwen-team-2024-qwen25}. Each query has three draws and five references
per draw. Generator robustness uses Mistral-7B-Instruct-v0.3
\citep{jiang-etal-2023-mistral} on up to 100 hash-selected queries from FiQA,
ArguAna, Touch\'e-2020, and SCIDOCS.

Dense retrieval uses BGE-small-en-v1.5 \citep{xiao-etal-2024-cpack}; normalized
query and document vectors are ranked by dot product. Sparse retrieval uses
Lucene BM25 through Pyserini 2.3.0, with $k_1=0.9$ and $b=0.4$
\citep{robertson-zaragoza-2009-bm25,lin-etal-2021-pyserini}. Contriever
\citep{izacard-etal-2022-contriever} is the second dense encoder. Document
identifiers break score ties.

\subsection{Comparison Methods}

Original submits the unexpanded query to both channels. Shared expansion holds
DESA's generated evidence fixed but applies it directly: Dense averages
the five query--reference representations, while Sparse concatenates the query
and references. Their comparison isolates channel-specific integration.

We reproduce three expansion baselines on all seven datasets: HyDE encodes
hypothetical documents for Dense \citep{gao-etal-2023-precise}; Query2doc
appends a pseudo-document to the query \citep{wang-etal-2023-query2doc}; and
MuGI pools reference representations for Dense and uses query-weighted
concatenation for Sparse \citep{zhang-etal-2024-exploring-best}.

The matched QuDAR control combines original-sparse (OS), original-dense (OD),
expanded-sparse (ES), and expanded-dense (ED) rankings
\citep{kim-etal-2026-qudar}. The main comparison fuses all four complete
rankings with uniform RRF ($\kappa=60$) and four-channel replay. The appendix reports
the official-style Top-1000 reproduction and a diagnostic confidence-weighted
variant.

\subsection{Mechanism and Robustness Analysis}

A $2\times2$ comparison covers Original, each operator alone, and DESA.
Matched controls isolate the operators. For Dense, the Original sparse ranking
is paired with unprojected contextual or orthogonal residual expansion. For
Sparse, the Original dense ranking is paired with direct rewriting or
score-product anchoring.

Dense behavior is measured by the angle between original and expanded query
vectors. Sparse behavior is measured by top-20 turnover and relevant-document
rank movement. Within-dataset quartiles and Spearman correlations relate these
quantities to effectiveness and certification depth.

Further controls vary the reference count over $\{1,3,5\}$, compare
score-product anchoring with a support-preserving binary mask, and use generated
passages alone for Sparse. Robustness analyses vary
$\kappa\in\{2,20,60,100\}$, the generator, the dense encoder, and corpus size.

\subsection{Statistical Analysis}

We average the three draws within each query before aggregation and report both
per-dataset results and equal-dataset macro averages. The 95\% confidence
intervals use 10,000 stratified bootstrap samples, resampling queries within
each dataset.

The primary comparisons are DESA against Original and Shared expansion. We use
10,000 stratified paired sign-flip randomizations and apply Holm correction
across nDCG@10, Recall@20, $L_D$, and $L_S$ for each comparator. Operator
controls use the same draw-averaged, query-level procedure. For the
post-held-out QuDAR controls, we use the same paired test. Holm correction is
applied separately to the two effectiveness metrics and the two total-depth
comparisons; total-depth reductions also receive bootstrap intervals.

\section{Results}
\label{sec:results}

\subsection{Fixed Cutoffs Alter Measured Effects}

\begin{figure}[H]
\centering
\includegraphics[width=\columnwidth]{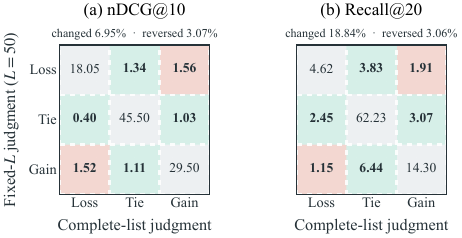}
\caption{Transition between the DESA-versus-Original judgment at fixed
$L=50$ (rows) and under complete-list fusion (columns). Cells report the
equal-dataset percentage of query--draw pairs. Green off-diagonal cells change
among gain, tie, and loss; red cells are strict gain--loss reversals.}
\label{fig:fixed-l-diagnostics}
\end{figure}

Figure~\ref{fig:fixed-l-diagnostics} compares DESA with Original at $L=50$ and
under complete-list fusion. Each query--draw pair is a gain, tie, or loss;
off-diagonal cells mark a changed judgment.

At $L=50$, 6.95\% of nDCG judgments and 18.84\% of Recall judgments differ
from complete-list fusion. Strict gain--loss reversals account for 3.07\% and
3.06\%, respectively. Other cutoffs and ordered top-20 agreement appear in
Appendix Table~\ref{tab:cutoff-query-diagnostics}.

\subsection{Complete-List Effectiveness and Certification Depth}

Table~\ref{tab:main-results} reports the primary complete-list comparison.
DESA reaches .4747 nDCG@10 and .4695 Recall@20, gains of 3.82\% and 2.38\%
over Original. Dense and sparse certification depths fall by 36.90\% and
36.56\%, and 63.31\% of queries become shallower in both channels. All four
differences remain significant after Holm correction
($p_{\mathrm{Holm}}=.0004$).

\begin{table}[H]
\centering
\scriptsize
\setlength{\tabcolsep}{3.2pt}
\begin{tabular}{lrrrr}
\toprule
Method & nDCG@10 $\uparrow$ & Recall@20 $\uparrow$ & $L_D\downarrow$ & $L_S\downarrow$ \\
\midrule
\multicolumn{5}{l}{\textit{Original query}} \\
Original & 0.4572 & 0.4586 & 1752.4 & 1455.1 \\
\addlinespace[1pt]
\multicolumn{5}{l}{\textit{Shared expansion}} \\
Shared & 0.4671 & 0.4666 & \textbf{607.4} & \textbf{606.7} \\
\addlinespace[1pt]
\multicolumn{5}{l}{\textit{Single-channel construction}} \\
Dense only  & 0.4621 & 0.4637 & 1386.5 & 1219.9 \\
Sparse only & \underline{0.4714} & \underline{0.4678} & 884.4 & 822.0 \\
\addlinespace[1pt]
\multicolumn{5}{l}{\textit{Channel-asymmetric construction}} \\
DESA & \textbf{0.4747} & \textbf{0.4695} & \underline{728.9} & \underline{673.2} \\
\bottomrule
\end{tabular}
\caption{Equal-dataset macro results for the seven-dataset controlled
comparison. Higher is better for retrieval metrics; lower is better for replay
stopping depths. Raw depths are averaged across datasets; percentage reductions
in the text are computed within each dataset before macro-averaging.}
\label{tab:main-results}
\end{table}

Shared expansion reaches lower raw depths but also lower nDCG. Because it uses
the same references as DESA, the difference isolates how shared evidence is
integrated across channels.

\subsection{Cross-Channel Coordination}

The matched Shared comparison isolates integration while holding generated
evidence fixed. DESA improves macro nDCG by 1.61\% and Recall by 0.61\%.
The nDCG difference is significant ($+.0075$;
$p_{\mathrm{Holm}}=.0012$), whereas Recall ($+.0028$;
$p_{\mathrm{Holm}}=.2736$) and both depth differences are not. The evidence
therefore supports a quality benefit from coordinated integration, but not a
depth advantage over Shared.

The sparse channel shows why shared evidence requires coordinated integration.
Shared expansion enlarges sparse support on six datasets, by as much as
$5.72\times$. On NFCorpus, support grows from 561.5 to 3211.6 documents and
sparse depth rises by 105.81\%. DESA instead preserves virtually all Original
support while reducing sparse and dense depth by 23.15\% and 15.03\%.
Table~\ref{tab:channel-evidence} extends this contrast across datasets.

\begingroup
\definecolor{effhead}{HTML}{DCEFEA}
\definecolor{accesshead}{HTML}{E2E7F5}
\definecolor{supporthead}{HTML}{FAEDC9}
\definecolor{positive}{HTML}{E7F5F0}
\definecolor{negative}{HTML}{F3D4CE}
\definecolor{access}{HTML}{EBEEF9}
\definecolor{broaden}{HTML}{FFF2CF}
\definecolor{preserve}{HTML}{DDF2E9}
\definecolor{pattern}{HTML}{EEF0F2}
\newcommand{\effcell}[1]{\colorbox{positive}{\strut\makebox[3.25em][c]{#1}}}
\newcommand{\negcell}[1]{\colorbox{negative}{\strut\makebox[3.25em][c]{\textbf{#1}}}}
\newcommand{\accesscell}[1]{\colorbox{access}{\strut\makebox[3.25em][c]{#1}}}
\newcommand{\broadcell}[1]{\colorbox{broaden}{\strut\makebox[3.65em][c]{#1}}}
\newcommand{\keepcell}[1]{\colorbox{preserve}{\strut\makebox[3.65em][c]{#1}}}
\newcommand{\patterncell}[1]{\colorbox{pattern}{\strut\scriptsize\textbf{#1}}}
\begin{table*}[t!]
\centering
\scriptsize
\setlength{\tabcolsep}{3.0pt}
\renewcommand{\arraystretch}{1.20}
\begin{tabular}{lrrrrrrr}
\toprule
& \multicolumn{3}{c}{\colorbox{effhead}{\strut\textbf{Effectiveness: relative $\Delta$nDCG@10 (\%)}}}
& \multicolumn{2}{c}{\colorbox{accesshead}{\strut\textbf{Replay-depth reduction (\%)}}}
& \multicolumn{2}{c}{\colorbox{supporthead}{\strut\textbf{Sparse support ($\times$ Original)}}} \\
\cmidrule(lr){2-4}\cmidrule(lr){5-6}\cmidrule(lr){7-8}
\textbf{Dataset}
& \textbf{DESA vs. Orig.}
& \textbf{+Dense given Sparse}
& \textbf{+Sparse given Dense}
& $\boldsymbol{L_D}$
& $\boldsymbol{L_S}$
& \textbf{Shared}
& \textbf{DESA} \\
\midrule
SciFact
& \effcell{+2.40} & \effcell{+1.12} & \effcell{+0.91}
& \accesscell{22.2} & \accesscell{20.9}
& \broadcell{1.62$\times$} & \keepcell{1.00$\times$} \\
NFCorpus
& \effcell{+3.37} & \effcell{+2.14} & \effcell{+1.36}
& \accesscell{15.0} & \accesscell{23.2}
& \broadcell{\textbf{5.72$\times$}} & \keepcell{1.00$\times$} \\
TREC-COVID
& \effcell{+5.69} & \effcell{+0.51} & \effcell{+4.68}
& \accesscell{74.9} & \accesscell{66.7}
& \broadcell{1.53$\times$} & \keepcell{1.00$\times$} \\
FiQA
& \effcell{+3.61} & \negcell{$-$0.66} & \effcell{+3.74}
& \accesscell{49.7} & \accesscell{48.7}
& \broadcell{1.64$\times$} & \keepcell{1.00$\times$} \\
ArguAna
& \effcell{+1.27} & \effcell{+0.32} & \effcell{+0.88}
& \accesscell{15.2} & \accesscell{15.3}
& \broadcell{1.00$\times$} & \keepcell{1.00$\times$} \\
Touch\'e-2020
& \effcell{+7.52} & \effcell{+0.92} & \effcell{+5.70}
& \accesscell{45.8} & \accesscell{45.8}
& \broadcell{1.88$\times$} & \keepcell{1.00$\times$} \\
SCIDOCS
& \effcell{+0.86} & \effcell{+0.03} & \effcell{+0.13}
& \accesscell{35.4} & \accesscell{35.4}
& \broadcell{2.02$\times$} & \keepcell{1.00$\times$} \\
\midrule
\textbf{Pattern}
& \patterncell{7/7 improve}
& \patterncell{6/7 improve}
& \patterncell{7/7 improve}
& \patterncell{7/7 shallower}
& \patterncell{7/7 shallower}
& \patterncell{up to 5.72$\times$}
& \patterncell{$\approx$1.00$\times$} \\
\bottomrule
\end{tabular}
\caption{Per-dataset evidence for cross-channel coordination.
Effectiveness columns report relative nDCG@10 changes; +Dense and +Sparse
compare DESA with the corresponding one-operator construction. Depth columns
report replay stopping-depth reductions from Original, where positive values
indicate smaller certified prefixes. Sparse support is normalized by Original;
red marks the sole degradation.}
\label{tab:channel-evidence}
\end{table*}
\endgroup

The $2\times2$ comparison reveals unequal but complementary roles. Dense
expansion alone improves nDCG and Recall by 1.08\% and 1.11\%, with dense and
sparse depth reductions of 8.16\% and 7.49\%. Sparse anchoring yields gains of
3.11\% and 2.01\%, with reductions of 28.77\% and 27.74\%. Adding Sparse
anchoring to Dense only improves nDCG on all seven datasets; adding Dense
expansion to Sparse only improves six, with FiQA at $-0.66\%$. Their
combination nevertheless improves over Original and reduces both depths on
every dataset.

The matched operator controls sharpen this conclusion. Residual dense expansion
does not significantly outperform unprojected expansion, nor does score-product
anchoring significantly outperform direct sparse rewriting. Yet full DESA
improves Original on every dataset and significantly improves nDCG over Shared.
The consistent advantage therefore belongs to the coordinated construction,
not to a demonstrably superior isolated operator. Sparse-support preservation
is exact on six datasets and 99.9999\% on Touch\'e-2020, whose exported tail
omits 25 extremely low-scoring document occurrences. Angle, turnover, rank-movement,
binary-mask, reference-count, and operator-control diagnostics appear in
Appendices~\ref{sec:appendix-mechanism} and~\ref{sec:appendix-analysis}.

\subsection{Comparison with Expansion and Fusion Methods}

Table~\ref{tab:fidelity} compares DESA with matched QuDAR and reproduced
expansion methods. Under matched references, QuDAR and DESA achieve similar
complete-list effectiveness; neither paired metric difference is significant.
DESA reduces total certification depth by 74.41\% (95\% CI: 67.27--80.89;
$p_{\mathrm{Holm}}=.0002$).

\begin{table}[H]
\centering
\scriptsize
\setlength{\tabcolsep}{2.4pt}
\begin{tabular}{lrrr}
\toprule
Method & nDCG@10 $\uparrow$ & Recall@20 $\uparrow$ & Total depth $\downarrow$ \\
\midrule
\multicolumn{4}{l}{\textit{Matched references, seven datasets}} \\
QuDAR-complete & \textbf{0.4760} & \textbf{0.4712} & \underline{5478.4} \\
DESA         & \underline{0.4747} & \underline{0.4695} & \textbf{1402.1} \\
\addlinespace[1pt]
\multicolumn{4}{l}{\textit{Prior query expansion, seven datasets}} \\
Original     & 0.4572 & 0.4586 & 3207.6 \\
HyDE         & 0.4114 & 0.4374 & 16668.8 \\
MuGI         & 0.4584 & 0.4596 & 2790.1 \\
Query2doc    & \underline{0.4728} & \underline{0.4675} & \underline{1771.5} \\
DESA         & \textbf{0.4747} & \textbf{0.4695} & \textbf{1402.1} \\
\bottomrule
\end{tabular}
\caption{External comparisons over the same seven evaluation datasets. Total
depth sums the policy-specific replay depths of all rankings used by each
complete-list fusion target.}
\label{tab:fidelity}
\end{table}

Among reproduced expansion methods, DESA has the highest effectiveness and the
lowest total depth. Its closest baseline, Query2doc, is slightly lower on both
metrics and requires greater total depth. These comparisons are descriptive;
we do not claim paired significance for the lower block of
Table~\ref{tab:fidelity}.

\subsection{Robustness and Failure Boundary}

With Mistral, all four sampled datasets retain non-negative effectiveness
changes. Dense and sparse depth reductions range from 7.12\% to 47.61\% and
7.16\% to 47.62\%; 10\% of sampled ArguAna queries trigger the structured-output
fallback.

With Contriever, both effectiveness metrics improve on all four datasets.
Touch\'e-2020 nevertheless increases dense and sparse depths by 14.29\% and
4.34\%, the only setting where both worsen. TREC-COVID gains remain positive
but non-monotonic across corpus sizes. Appendix~\ref{sec:additional-results}
reports the remaining generator, encoder, reference-count, fusion, and scale
results.

\FloatBarrier

\section{Conclusion}
\label{sec:conclusion}

This work frames hybrid query expansion as a coupled construction problem for
the fused result. A fixed top-$L$ cutoff entangles expansion effects with the
prefixes used for fusion, so we separate complete-list effectiveness from the
per-channel depths needed to certify the same top-$K$. DESA designs both
channels for this joint objective using shared generated references with
channel-specific evidence authority: orthogonal residual expansion adds new
dense semantic directions, while score-product anchoring reorders only the
original sparse support.

Across seven BEIR datasets, DESA improves equal-dataset macro nDCG@10 and
Recall@20 over Original by 3.82\% and 2.38\%, while reducing dense and sparse
replay stopping depths by 36.90\% and 36.56\%. Under matched generated
references, it also provides a significant nDCG advantage over Shared expansion
and preserves the original sparse support. Against a matched complete-list
four-signal QuDAR control, the effectiveness difference is statistically
unresolved, while DESA requires 74.41\% less total certification depth. Sparse
anchoring supplies the more consistent effectiveness gains, while dense
expansion contributes additional gains on most datasets; matched controls do
not establish either exact operator as uniquely optimal.

Because the fused target depends jointly on both rankings, generated evidence
must be coordinated across them. It may extend dense semantics without
acquiring independent sparse admission authority: Dense expands; Sparse anchors.

\section*{Limitations}

Our evaluation covers seven English BEIR datasets, two generators, two dense
encoders, BM25, equal-weight RRF, and an ordered top-20 target. Generalization
to other languages, learned sparse retrievers, or fusion rules remains
untested, and score-product anchoring assumes non-negative sparse scores.

Replay stopping depth counts ranked evidence under the fixed policy; it is not
latency, compute, or physical index work. End-to-end behavior also depends on
generation, encoding, indexing, and incremental ranking access. The matched
QuDAR comparison extends the same replay principle to four rankings, but its
total depth should still be interpreted only as policy-specific logical access.

Sparse anchoring excludes expansion-only matches, which limits lexical drift
but may suppress relevant documents reachable only through generated terms.
Generated references and pooled judgments may also be unreliable or
incomplete. Certification depth increases with Contriever on Touch\'e-2020,
and scale effects are non-monotonic. Future work should test selective support
relaxation and online ranked access.

\bibliography{references}

\appendix
\section{Reproducibility Details}
\label{sec:appendix-reproducibility}
\raggedbottom
\makeatletter
\setlength{\@fptop}{0pt}
\setlength{\@fpsep}{12pt plus 2pt minus 2pt}
\setlength{\@fpbot}{0pt plus 1fil}
\setlength{\@dblfptop}{0pt}
\setlength{\@dblfpsep}{12pt plus 2pt minus 2pt}
\setlength{\@dblfpbot}{0pt plus 1fil}
\makeatother

This section records the generation, query-construction, retrieval, and replay
settings needed to reproduce the experiments.

\subsection{Complementary Reference Generation}

We use the following Qwen2.5-7B-Instruct revision for both the model and
tokenizer:

{\footnotesize\noindent\texttt{fe11104b620d588ccc049ff6631dd3ea002e3d98}\par}

The unquantized BF16 model runs with MLX-LM 0.31.2. Decoding uses a
temperature of 0.7, top-$p$ of 0.9, top-$k$ of 20, and a maximum of 256 new
tokens. Queries are sorted by identifier and processed in batches of eight.
Each batch seed is derived from the protocol version, dataset, prompt hash, and
batch index using the first 64 bits of their SHA-256 digest.

\begin{promptbox}{Prompt 1. Complementary Reference Generation}
\textbf{System prompt}

You generate evidence-like reference passages for information retrieval.

Given one search query, return exactly five concise, complementary passages
that could plausibly occur in relevant documents. Preserve every named entity,
relation, number, polarity, temporal condition, scope, and constraint in the
query. Add useful aliases, technical terminology, and paraphrases, but do not
broaden the information need or invent a specific answer. Each passage must be
one sentence of at most 25 words.

Return one single-line JSON array containing exactly five distinct strings.
Count the five array items before responding. Do not output an object key,
query identifier, Markdown, placeholders, or any other text. Every item must
contain a substantive passage.

\par\smallskip\hrule\smallskip
\textbf{User input}\par
\texttt{\{"query":"\{query\}"\}}
\end{promptbox}

The user input is serialized as a canonical JSON object with one
\texttt{query} field.
XGrammar 0.2.1 requires exactly five strings; validation rejects empty,
duplicate, or placeholder passages. Invalid output is retried once with the
same seed and validation feedback, then falls back to the unexpanded query.

The second generator is Mistral-7B-Instruct-v0.3 at revision

{\footnotesize\noindent\texttt{c170c708c41dac9275d15a8fff4eca08d52bab71}.\par}

It uses the same decoding, batching, and validation settings.

\subsection{Query Construction}

Orthogonal residual expansion uses float32 throughout. Shared expansion uses
the same \texttt{query [SEP] reference} representations, but averages and
normalizes them without removing the component parallel to the original query.

For sparse retrieval, the expanded query contains one copy of the original
query followed by the selected reference passages. BM25 scores under the
original and expanded queries are stored separately before applying the
document-wise score product. The binary-mask ablation retains the expanded-query
score only for documents matched by both queries; the references-only ablation
removes the original query. DESA neither repeats the original query nor uses
MuGI's repetition coefficient.

\subsection{Baseline Reproduction}

All generative baselines use the same primary generator. Their generation and
query-integration settings are specified below.

HyDE generates eight hypothetical documents per run and averages their
normalized representations; its sparse channel uses the original query. FiQA,
ArguAna, SciFact, and TREC-COVID use task-specific instructions, the remaining
datasets use a general instruction, and generation is capped at 512 new tokens.

Query2doc generates one pseudo-document of at most 120 words. Dense retrieval
encodes it with the original query; sparse retrieval concatenates it with five
copies of the original query.

MuGI generates five pseudo-references. Dense retrieval averages the five
query--reference representations. Sparse retrieval repeats the original query

\[
\left\lfloor\frac{\sum_{i=1}^{n}|r_i|}{|q_o|\beta}\right\rfloor
\]

times before appending all references, where $|\cdot|$ denotes character
length and $\beta=4$. We reproduce MuGI's generation and query integration
components without its pseudo-relevance feedback calibration.

\subsection{Retrieval Implementation}

BGE-small-en-v1.5 uses revision

{\footnotesize\noindent\texttt{5c38ec7c405ec4b44b94cc5a9bb96e735b38267a}.\par}

Queries are prefixed with
``Represent this sentence for searching relevant passages:'' and encoded
with a maximum length of 512 tokens. Contriever uses mean pooling at revision

{\footnotesize\noindent\texttt{2bd46a25019aeea091fd42d1f0fd4801675cf699}.\par}

All query and document embeddings are normalized. Pyserini 2.3.0 runs with
Java 21 and uses \texttt{DefaultEnglishAnalyzer}. A separate Lucene index is
built for each dataset and each TREC-COVID snapshot, so BM25 collection
statistics are computed within the corresponding corpus. Documents with equal
retrieval or fusion scores are ordered by the UTF-8 order of their identifiers.

\subsection{Access-Depth Replay}

This section gives implementation details for the bound-based replay defined
in Section~\ref{sec:quality-access}. The replay starts with empty dense and
sparse prefixes. After reading $l_c$
items from channel $c$, its next unread contribution is bounded by

\[
u_c(l_c)=\frac{1}{60+l_c+1}.
\]

The bound becomes zero after exhaustion. The replay reads the channel with the
larger bound, breaking ties toward dense. An observed document's accumulated
score is a lower bound; an unseen contribution is bounded by the corresponding
next-unread value, and a document unseen in both channels by their sum. Replay
stops when these bounds certify top-20 membership and order, recording
$(L_D,L_S)=(l_D,l_S)$.

\subsection{Artifact Traceability}

Generation records store the raw and parsed output, hashes, decoding settings,
seed, retries, runtime lock, and code commit. Query-level records store the
method, generation run, effectiveness, $(L_D,L_S)$, sparse support and
exhaustion, ordered top-20, and hashes of the compressed rankings,
configuration, generation, and replay trace.

The Python environment is fixed by \texttt{uv.lock}; Java is checked before
retrieval. A hash manifest records code, configurations, prompts, inputs,
models, generations, and rankings. Reported tables are rebuilt from the
query-level records.

\section{Operator Properties and Mechanism Diagnostics}
\label{sec:appendix-operators}

\subsection{Properties of the Channel Operators}

For the dense construction, each contextual vector $\mathbf v_i$ is unit
normalized. The triangle inequality gives

\[
\|\bar{\mathbf v}_r\|
\leq \frac{1}{n}\sum_{i=1}^{n}\|\mathbf v_i\|=1.
\]

Orthogonal projection cannot increase the norm, so
$\|\mathbf z\|\leq 1$. Since $\mathbf v_o^\top\mathbf z=0$ and
$\|\mathbf v_o\|=1$,

\[
\begin{aligned}
\mathbf e_D^\top\mathbf v_o
&=\frac{1}{\sqrt{1+\|\mathbf z\|^2}},\\
\angle(\mathbf e_D,\mathbf v_o)
&=\arctan\|\mathbf z\|\leq\frac{\pi}{4}.
\end{aligned}
\]

If $\mathbf z=\mathbf 0$, normalization leaves $\mathbf v_o$ unchanged.

For sparse retrieval, non-negativity gives
$\operatorname{supp}(s_os_r)=
\operatorname{supp}(s_o)\cap\operatorname{supp}(s_r)$. Additivity and the
inclusion of $q_o$ in $q_r$ imply
$\operatorname{supp}(s_o)\subseteq\operatorname{supp}(s_r)$, yielding exact
support preservation. If $s_r=c\,s_o$ for $c>0$, then
$s_A=c\,s_o^2$; squaring is strictly increasing on positive scores, so the
ranking is unchanged. Multiplying $s_o$ or $s_r$ by a positive global constant
only rescales every anchored score and likewise leaves the ranking unchanged.

\subsection{Mechanism Diagnostics}
\label{sec:appendix-mechanism}

Table~\ref{tab:mechanism-diagnostics} reports dense angle, one minus sparse
top-20 overlap with Original, and the mean reciprocal-rank change of judged
relevant documents present in the exported rankings. Macros weight datasets
equally.

\begin{table*}[!t]
\centering
\small
\setlength{\tabcolsep}{5.0pt}
\begin{tabular}{lrrrr}
\toprule
Dataset & Dense angle ($^\circ$) & Sparse top-20 turnover (\%) & Relevant RR change & Support retained (\%) \\
\midrule
SciFact        &  9.38 & 18.88 & .0211 & 100.0000 \\
NFCorpus       & 17.65 & 18.69 & .0147 & 100.0000 \\
TREC-COVID     & 12.99 & 45.37 & .0021 & 100.0000 \\
FiQA           & 13.34 & 33.91 & .0204 & 100.0000 \\
ArguAna        &  4.02 &  8.14 & .0048 & 100.0000 \\
Touch\'e-2020 & 14.40 & 37.07 & .0091 &  99.9999 \\
SCIDOCS        & 12.05 & 20.98 & .0028 & 100.0000 \\
\midrule
Equal-dataset macro & 11.98 & 26.15 & .0107 & 100.0000 \\
\bottomrule
\end{tabular}
\caption{Mechanism diagnostics computed from the frozen query constructions
and rankings. Support retained is the mean per-query--draw fraction of the
original sparse support; the displayed macro rounds to 100.0000\%.}
\label{tab:mechanism-diagnostics}
\end{table*}

\begin{figure}[H]
\centering
\includegraphics[width=\columnwidth]{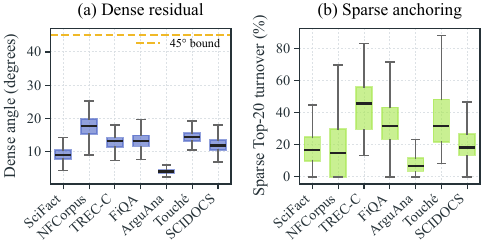}
\caption{Realized behavior of the two channel operators over three generation
draws on seven datasets. Dense angles are measured from the original query;
sparse turnover is one minus top-20 overlap with the original sparse ranking.
Boxes show the interquartile range and median; whiskers extend to 1.5 times the
interquartile range, with outliers omitted.}
\label{fig:mechanism-distributions}
\end{figure}

Figure~\ref{fig:mechanism-distributions} shows the query-level distributions
behind the aggregate diagnostics. One invalid-generation fallback leaves
11,327 query--draw pairs. Their mean
dense angle is $11.98^\circ$ and the maximum is $27.87^\circ$, below the
analytic $45^\circ$ bound. Sparse anchoring changes 26.15\% of the original
sparse top-20 on average.

Within-dataset quartiles show no monotonic relation between either diagnostic
and effectiveness or access depth. Relevant-document reciprocal-rank gain
correlates more strongly with sparse-only nDCG gain than overall turnover does
(.391 versus .078); dense-angle correlations with nDCG gain and depth reduction
are .010 and .011. The data therefore support no angle or turnover threshold.

Exact support preservation assumes a non-negative additive scorer and complete
score export. It holds for six datasets. Touch\'e-2020 omits 25 extremely
low-scoring tail documents across 23 of 147 query--draw pairs; mean and minimum
retention are 99.99990\% and 99.99881\%. We retain these cases and do not call
the Lucene export exactly support preserving there.

\section{Additional Experimental Results}
\label{sec:appendix-analysis}

\subsection{Additional Ablations and Robustness}
\label{sec:additional-results}

Binary masking preserves sparse support but removes original-score magnitude,
lowering macro nDCG@10 by 1.54\% and Recall@20 by 0.74\%. Increasing references
from one to three improves nDCG@10 by 1.91\% and reduces dense/sparse replay
stopping depths by 17.23\%/18.14\%; using five adds 0.35\% nDCG@10 and 0.43\%
Recall@20, with further reductions of 3.83\%/3.40\%.

With Mistral, all four sampled datasets retain non-negative effectiveness
changes and 7.12--47.61\%/7.16--47.62\% dense/sparse depth reductions; 10\% of
sampled ArguAna queries fall back. Contriever improves both metrics on all four
datasets but raises Touch\'e-2020 depths by 14.29\%/4.34\%. Nested TREC-COVID
gains and reductions remain positive but non-monotonic.

\subsection{Matched Operator and QuDAR Controls}

Table~\ref{tab:operator-controls} reports matched operator controls. Operator
pairs share references, keep the other channel at Original, and average three
draws within query before inference. The QuDAR control uses the same generated
references as DESA and complete-list RRF over OS, OD, ES, and ED.

\begin{table*}[!t]
\centering
\footnotesize
\textbf{(a) Per-dataset controlled results}\par\smallskip
\setlength{\tabcolsep}{2.2pt}
\begin{tabular}{llrrrrrrr}
\toprule
Dataset & Method & Queries & nDCG@10 & Recall@20 & $L_D$ & $L_S$ & Sparse support & Exhaust. (\%) \\
\midrule
ArguAna & Original & 1406 & .3941 & .9203 & 157.1 & 156.7 & 8587.8 & 0.00 \\
        & Shared   &      & .4004 & .9265 & 123.4 & 123.0 & 8616.8 & 0.00 \\
        & DESA     &      & .3991 & .9237 & 133.1 & 132.7 & 8587.8 & 0.00 \\
\addlinespace
FiQA & Original & 648 & .3649 & .5260 & 2179.0 & 2076.3 & 28384.6 & 1.23 \\
     & Shared   &     & .3635 & .5391 & 1104.1 & 1103.6 & 46673.3 & 0.00 \\
     & DESA     &     & .3780 & .5450 & 1096.2 & 1065.6 & 28384.6 & 0.93 \\
\addlinespace
NFCorpus & Original & 323 & .3670 & .2099 & 474.4 & 173.7 & 561.5 & 48.61 \\
         & Shared   &     & .3780 & .2260 & 359.1 & 357.6 & 3211.6 & 1.14 \\
         & DESA     &     & .3793 & .2195 & 403.1 & 133.5 & 561.5 & 46.75 \\
\addlinespace
SCIDOCS & Original & 1000 & .1923 & .2721 & 694.8 & 599.3 & 11503.0 & 2.70 \\
        & Shared   &      & .1905 & .2790 & 372.7 & 372.2 & 23239.6 & 0.00 \\
        & DESA     &      & .1940 & .2787 & 448.8 & 387.2 & 11503.0 & 1.63 \\
\addlinespace
SciFact & Original & 300 & .7253 & .9036 & 409.1 & 368.6 & 2765.7 & 4.33 \\
        & Shared   &     & .7325 & .9182 & 284.1 & 282.6 & 4489.0 & 0.33 \\
        & DESA     &     & .7427 & .9200 & 318.3 & 291.6 & 2765.7 & 3.89 \\
\addlinespace
TREC-COVID & Original & 50 & .7781 & .0398 & 6270.9 & 4729.9 & 92683.8 & 2.00 \\
           & Shared   &    & .8169 & .0433 & 787.3 & 786.9 & 141483.1 & 0.00 \\
           & DESA     &    & .8224 & .0435 & 1574.8 & 1574.4 & 92683.8 & 0.00 \\
\addlinespace
Touch\'e-2020 & Original & 49 & .3786 & .3382 & 2081.8 & 2081.4 & 151619.0 & 0.00 \\
              & Shared   &    & .3883 & .3344 & 1221.5 & 1221.3 & 284677.0 & 0.00 \\
              & DESA     &    & .4071 & .3560 & 1127.8 & 1127.4 & 151618.9 & 0.00 \\
\bottomrule
\end{tabular}

\medskip
\begin{minipage}[t]{0.49\textwidth}
\centering
\textbf{(b) Primary statistical tests}\par\smallskip
\setlength{\tabcolsep}{1.2pt}
\begin{tabular}{lll}
\toprule
Comparison / outcome & Effect [95\% CI] & $p/p_{\mathrm{Holm}}$ \\
\midrule
Original / nDCG@10   & .0175 [.0111, .0238]    & .0001 / .0004 \\
Original / Recall@20 & .0109 [.0070, .0151]    & .0001 / .0004 \\
Original / $L_D$     & 1023.6 [377.5, 2012.6]  & .0001 / .0004 \\
Original / $L_S$     & 781.9 [356.1, 1339.7]   & .0001 / .0004 \\
\addlinespace
Shared / nDCG@10     & .0075 [.0036, .0116]     & .0003 / .0012 \\
Shared / Recall@20   & .0028 [-.0003, .0061]    & .0912 / .2736 \\
Shared / $L_D$       & -121.4 [-367.0, 39.7]    & .3157 / .6313 \\
Shared / $L_S$       & -66.5 [-310.6, 93.0]     & .7909 / .7909 \\
\bottomrule
\end{tabular}
\end{minipage}\hfill
\begin{minipage}[t]{0.49\textwidth}
\centering
\textbf{(c) Fixed-cutoff diagnostics}\par\smallskip
\setlength{\tabcolsep}{2.2pt}
\begin{tabular}{rrrrrr}
\toprule
$L$ & Exact top-20 & \multicolumn{2}{c}{nDCG (\%)} & \multicolumn{2}{c}{Recall (\%)} \\
 & (\%) & Change & Reverse & Change & Reverse \\
\midrule
20   & 1.92 & 18.23 & 8.84 & 22.50 & 3.84 \\
50   & 16.54 & 6.95 & 3.07 & 18.84 & 3.06 \\
100  & 47.40 & 2.17 & 0.73 & 11.74 & 0.99 \\
1000 & 92.62 & 0.19 & 0.01 & 1.18 & 0.00 \\
\bottomrule
\end{tabular}
\end{minipage}
\caption{Additional results. (a) Dataset-level controlled results. (b)
Stratified paired tests after within-query averaging over three generation
draws. (c) DESA query--draw sensitivity to fixed top-$L$, macro-averaged with
equal dataset weight; exact top-20 compares with complete-list fusion, and the
conclusion columns compare DESA with Original.}
\label{tab:per-dataset-controlled}
\label{tab:primary-tests}
\label{tab:cutoff-query-diagnostics}
\end{table*}

\begin{table}[H]
\centering
\scriptsize
\setlength{\tabcolsep}{2.8pt}
\begin{tabular}{@{}lrr@{}}
\toprule
Construction & nDCG@10 & Recall@20 \\
\midrule
Original          & .4572 & .4586 \\
Contextual Dense  & .4622 & .4637 \\
Residual Dense    & .4621 & .4637 \\
Rewrite Sparse    & .4686 & .4670 \\
Product Sparse    & .4714 & .4678 \\
\midrule
\multicolumn{3}{l}{Paired effect (proposed $-$ control)} \\
Dense residual, nDCG   & \multicolumn{2}{l}{$-.00005$ [$-.00086,.00082$], $p_{\mathrm{Holm}}=1.0$} \\
Dense residual, Recall & \multicolumn{2}{l}{$-.00006$ [$-.00043,.00033$], $p_{\mathrm{Holm}}=1.0$} \\
Sparse product, nDCG   & \multicolumn{2}{l}{$ .00283$ [$-.00126,.00693$], $p_{\mathrm{Holm}}=.3650$} \\
Sparse product, Recall & \multicolumn{2}{l}{$ .00078$ [$-.00184,.00331$], $p_{\mathrm{Holm}}=.5689$} \\
\bottomrule
\end{tabular}
\caption{Macro operator controls and query-level paired effects. Holm
correction covers nDCG and Recall within each control pair.}
\label{tab:operator-controls}
\end{table}

Against complete uniform QuDAR, DESA's paired effects are $-.00132$
[$-.00470,.00194$] for nDCG and $-.00174$ [$-.00382,.00033$] for Recall
($p_{\mathrm{Holm}}=.4393$ and $.2140$), while reducing total certification
depth by 74.41\% [67.27\%, 80.89\%] ($p_{\mathrm{Holm}}=.0002$).

\begingroup\footnotesize
\noindent\textbf{Confidence-weighting audit.}
Replacing QuDAR's margin weights with $w_c=.25$ while holding its inputs fixed
yields no significant nDCG/Recall difference
($p_{\mathrm{Holm}}=.1182/.5826$); 92.14\%/99.04\% of pooled query--draw
outcomes are unchanged, with dataset means in .2325--.2699. In a
complete-list RRF diagnostic, the weights lower nDCG by .00117
($p_{\mathrm{Holm}}=.0104$), leave Recall unchanged
($p_{\mathrm{Holm}}=.4312$), and raise depth from 5478.4 to 5601.4. Neither
test shows an independent benefit in this setting; the unofficial RRF transfer
does not assess LLM weighting or QuDAR overall.
\endgroup

\end{document}